# Simulation of Hydrogen Adsorption in Hierarchical Silicalite: Role of Electrostatics and Surface Chemistry

Siddharth Gautam[1*], David R. Cole[1], Zoltán Imre Dudás[2] and Indu Dhiman[2]

[1]School of Earth Sciences, The Ohio State University, 275 Mendenhall Laboratory, 125 S Oval Mall, Columbus, OH 43210, USA

[2]HUN-REN Centre for Energy Research, Neutron Spectroscopy Department, Budapest, 1121, Hungary

## Abstract

Adsorption in nanoporous materials is one strategy that can be used to store hydrogen at conditions of temperature and pressure that are economically viable. Adsorption capacity of nanoporous materials depends on surface area which can be enhanced by incorporating a hierarchical pore structure. We report grand canonical Monte Carlo (GCMC) simulation results on the adsorption of hydrogen in hierarchical models of silicalite that incorporate 4 nm wide mesopores in addition to the 0.5 nm wide micropores at 298 K, using different force fields to model hydrogen. Our results suggest that incorporating mesopores in silicalite can enhance adsorption by at least 20% if electrostatic interactions are not included and up to 100% otherwise. Incorporating electrostatic interactions results in higher adsorption by close to 100% at lower pressures for hierarchical silicalite whereas for unmodified silicalite, it is less significant at all pressures. Hydroxylating the mesopore surface in hierarchical silicalite results in an enhancement in adsorption at pressures below 1 atm and suppression by up to 20 % at higher pressures. Temperature dependence at selected pressures exhibits expected decrease in adsorption amounts at higher temperatures. These findings can be useful in the engineering, selection, and optimization of nanoporous materials for hydrogen storage.

*Corresponding author email: gautam.25@osu.edu

## 1.0 INTRODUCTION

Use of fossil fuels has given rise to a dramatic increase in the atmospheric content of greenhouse gases like $CO_2$ that in turn are responsible for global warming [1, 2]. To control this warming effect, several strategies have been proposed including the use of alternative and renewable energy resources that can replace fossil fuels [3]. Hydrogen is one such fuel that is promising due to its high energy density [4]. However, the thermodynamic properties of hydrogen [5] make it difficult to store this fluid [6]. For example, storing hydrogen in compressed gas containers requires very high pressures and can lead to concerns regarding safety in handling such highly compressed volatile fluids [7], in addition to the economic cost of compression [8]. Another possibility is to store it as a liquified gas, but since the boiling point of hydrogen is 20 K, it requires costly cryogenic facilities [9]. For these reasons, alternative cost-effective strategies for hydrogen storage are required that can reduce the economic cost of storage. Storing hydrogen by chemically reacting it with metals to form metal hydrides has been proposed as such a strategy [10]. Another strategy is physically adsorbing it in nanoporous materials [11]. Physical sorption is a safe and cost-effective strategy. For this reason, hydrogen adsorption in various nanoporous materials has been investigated [12 – 23]. Silica based nanoporous materials offer a wide range of pore morphology and surface area characteristics [24]. In particular silicalite, an all-silica analogue of ZSM-5 is a silica-based zeolite that has sub-nanometer sized channel like pores with high degree of inter-connectivity [25]. This results in a high surface area making this material suitable for gas storage applications [26]. Adsorption of several gases of industrial and environmental importance in silicalite has been studied and reported [26 – 32].

Nanoporous adsorbents suitable for gas storage are expected to have optimal porosity and surface area. In some cases, surface area of nanoporous materials can be enhanced by introducing a hierarchical pore structure [33, 34]. In such cases, nano-pores of different dimensions can co-exist in the same material. Silicalite has pores that are classified as micropores by the IUPAC classification [35]. Introducing mesopores in silicalite can give rise to a hierarchical structure with higher surface area. *Can this structure exhibit better hydrogen storage capability?* To address this question, we carried out a grand canonical Monte Carlo (GCMC) simulation study on hydrogen adsorption in silicalite and its hierarchical counterpart with 4 nm wide cylindrical pores etched in different directions.

Several force-fields for simulating hydrogen have been proposed [23]. Some of these account for electrostatic interactions while others ignore them. If the electrostatic interactions of hydrogen are switched off in a simulation, the electrostatic constitution of the adsorbent material is not expected to affect the adsorption behavior and hence nanoporous materials with the same pore properties but different surface chemistries (e.g., presence of hydroxyl groups) may exhibit similar adsorption behavior. *How justified can this similarity be?* To address this question, we also carry out some of our simulations with different force-fields accounting for a different treatment of electrostatic interactions.

Our simulations suggest that electrostatic interactions can significantly determine the adsorption amounts at room temperature. Also, existence of mesopores in addition to micropores lead to higher adsorption amounts of hydrogen. In section 2, we list the details of our simulations. This includes the building up of hierarchical silicalite structures, the force-fields used and the simulation method and related parameters. In section 3 we list the major results and discuss their implications in the light of published literature in section 4. Finally, we close with our concluding remarks in section 5.

## 2.0 SIMULATION DETAILS

### 2.1 Adsorbents

Silicalite is an all-silica analogue of ZSM-5 zeolite [25]. It has channel like straight pores approximately 0.55 nm in diameter running along the crystallographic *b* axis. These straight channels are connected to each other via zig-zag pores of similar size that run sinusoidally in the *a-c* plane. Atomic coordinates of silicalite in a unit cell as provided by Konigsveld et al [25] were used to make a larger supercell by replicating the unit cell 4×4×6 times in all directions. This resulted in a supercell of dimensions 7.95×8.04×8.02 $nm^3$ and constituted an ideal crystalline model of silicalite (henceforth referred to as IC). A cylindrical mesopore of diameter 4 nm was then carved out from this cell by removing atoms that lay within a cylindrical volume of diameter 4 nm and spanning the entire volume in the perpendicular direction. Atoms were removed such that the final configuration had Si and O atoms in 1:2 ratio to maintain the overall charge neutrality. A quarter cylinder was also carved out on each corner of the cell. Thus, each cell had effectively two mesopores (1+4(1/4)) (see Figure 1). Since silicalite is anisotropic (straight channels along *b* and sinusoidal channels in the *a-c* plane), three different models of mesoporous silica were made with (i) mesopores lying in the *c* direction (This hierarchical structure with two mesopores with their axes along the Cartesian Z direction is termed "HZZ-NH". The first letters here indicate the presence of pore hierarchy, next two indicate the direction of the two mesopores and the final two letters in the suffix stand for presence (OH) or absence (NH) of hydroxyl groups on the mesopore surface), (ii) mesopores lying in the *b* direction (termed "HYY-NH"), and (iii) mesopores lying perpendicular to each other and along the directions *b* and *c* (termed "HYZ-NH"). Further, as atoms are removed from the original silicalite simulation cell to generate mesoporous models, some atoms lying on the pore surface remain unbonded. To saturate their broken bonds, we added -OH groups to unsaturated Si atoms and H atoms to unsaturated O atoms, resulting in full saturation. Short MD simulations using DL_Poly [36] were then used to relax the newly added -OH groups. The pore surface was thus covered with -OH groups. To understand the effects of this hydroxylation, we simulated both the hydroxylated and non-hydroxylated versions of the structure HZZ with the corresponding hydroxylated version termed "HZZ-OH". We note that the hydroxylated version corresponds to a full hydroxylation with no system investigated with partial hydroxylation. The effects of hydroxylation are therefore investigated as binary absence or presence of these groups. A simulation snapshot showing the model HZZ-OH in the *a-b* and *a-c* planes is included in Figure 1 (a) while schematics of the other models are included in Figure 1 (b)

## 2.2 Adsorbate Models and Force-Fields

ClayFF force-field is often used to simulate mineral oxides including silica [37]. We have used it earlier to represent silicalite in several studies and in combination with TraPPE force-fields for ethane and $CO_2$, it gave a good agreement with the experimental data for adsorption of ethane and $CO_2$ in silicalite [31]. Several force-fields have been parametrized for modelling hydrogen and have their advantages/disadvantages in modelling the behavior of hydrogen. Since hydrogen is a light molecule, quantum effects become important to account for in modelling this gas especially at low temperatures (e.g. at 77 K at which many experimental hydrogen adsorption studies are carried out). However, if applications in hydrogen storage are considered, the relevant temperatures are close to room temperature and the quantum effects are no longer important even for hydrogen [38]. The simplest force-field for hydrogen is proposed by Buch and treats hydrogen molecule as a single structureless spherical entity that interacts with other species solely via van der Waals interaction that take the form of a Lennard-Jones potential [39]. This is a cheap force-field as it has no electrostatic interactions to model. However, since hydrogen molecules have a significant quadropole moment [40], they cannot be modelled as single spherical entities unless the quadrupolar moment is to be ignored. Therefore, to incorporate the quadrupole moment, Darkrim and Levesque proposed a force-field that treats hydrogen as a linear three-site molecule with two sites representing hydrogen atoms separated by 0.074 nm and a third site representing the center of mass kept at the center [38] (see Figure S1 in the supplementary information appended at the end of this document). The quadrupolar moment is accounted for by assigning partial charges to all the three sites. It has been shown, however, that accounting for electrostatic interactions results in overestimation of adsorbed amounts at some pressures, thus eliminating the charges could give better estimates of adsorption amounts [23]. For this reason, we utilized three force-fields here –(i) Buch, (ii) Darkrim-Levesque (DL) and (iii) Darkrim-Levesque, but with partial charges on the three sites removed (DLnc). All parameters related to all the force-fields mentioned above are provided in the supplementary information.

### 2.3 Simulations

Grand canonical Monte Carlo simulations were carried out using DL-Monte [41]. The starting configuration consisted of a single hydrogen molecule placed in a pore of any of the models mentioned earlier. Using this starting configuration Monte Carlo moves that consisted of a random translation, rotation (only for DL and DLnc force-fields), addition or deletion of an adsorbate molecule were attempted, subject to the given environmental conditions of temperature and partial pressure of the adsorbate (henceforth referred to as pressure for brevity). As the simulation progressed, the number of adsorbate molecules were monitored. This number exhibited a systematic variation up to 50,000 steps and achieved a stable value after that. The configurations obtained during these initial steps were therefore discarded and those obtained during subsequent 150,000 steps were used to obtain averages of equilibrium quantities. Simulations were carried out at 298 K and a total of 22 pressures for each system between 0.05 atm and 10 katm. We note that at 298 K, bulk hydrogen becomes supercritical at 16.7 atm and its density between 0.05 atm and 10 katm ranges between 4 micrograms/ml to 0.137 grams/ml [5] The lowest pressure was simulated first, and subsequent simulations were carried out at progressively higher pressures using the final configuration of the previous simulations. This resulted in achievement of equilibrium much sooner than 50,000 steps. However, for consistency, data in all simulations were averaged over 150,000 steps following the initial 50,000 steps. In addition to the simulations at 298 K, some simulations were also carried out at different temperatures between 273 and 308 K to study the effects of temperature on the adsorption amounts. In all, 332 simulations were carried out differing in pressures, adsorbent models, adsorbate model (force-field), chemistry of the pore surface (hydroxylated vs non-hydroxylated pore surface) and temperatures. Table 1 provides a summary of these simulations.

**Table 1.** Simulations reported in this work on hydrogen adsorption in silicalite models. Different models of silicalite are used and hydrogen is modeled using different force-fields. In all 332 simulations tabulated here were carried out at different temperature and pressure conditions noted.

| Adsorbent Model | Pore Surface Chemistry | Adsorbate Model (force-field) | Temperature (K) | Pressures | Number of simulations |
|---|---|---|---|---|---|
| Ideal silicalite (IC) | Non-hydroxylated | Buch | 298 | 0.05 atm – 10 katm | 22 |
| | | | 273; 278; 283; 288; 293; 303 and 308 | Each temperature at 0.1, 1.0, 10.0 and 100.0 atm | 28 |
| | | DL | 298 | 0.05 atm – 10 katm | 22 |
| | | | 273; 278; 283; 288; 293; 303 and 308 | Each temperature at 0.1, 1.0, 10.0 and 100.0 atm | 28 |
| | | DLnc | 298 | 0.05 atm – 10 katm | 22 |
| HZZ-NH *w/mesopore* | Non-hydroxylated | Buch | 298 | 0.05 atm – 10 katm | 22 |
| | | DL | 298 | 0.05 atm – 10 katm | 22 |
| HYY-NH *w/mesopore* | Non-hydroxylated | DL | 298 | 0.05 atm 10 katm | 22 |
| HYZ-NH *w/mesopore* | Non-hydroxylated | DL | 298 | 0.05 atm –10 katm | 22 |
| HZZ-OH *w/mesopore* | Hydroxylated | Buch | 298 | 0.05 atm –10 katm | 22 |
| | | | 273; 278; 283; 288; 293; 303 and 308 | Each temperature at 0.1, 1.0, 10.0 and 100 atm | 28 |
| | | DL | 298 | 0.05 atm – 10 katm | 22 |
| | | | 273; 278; 283; 288; 293; 303 and 308 | Each temperature at 0.1, 1.0, 10.0 and 100.0 atm | 28 |
| | | DLnc | 298 | 0.05 atm – 10 katm | 22 |

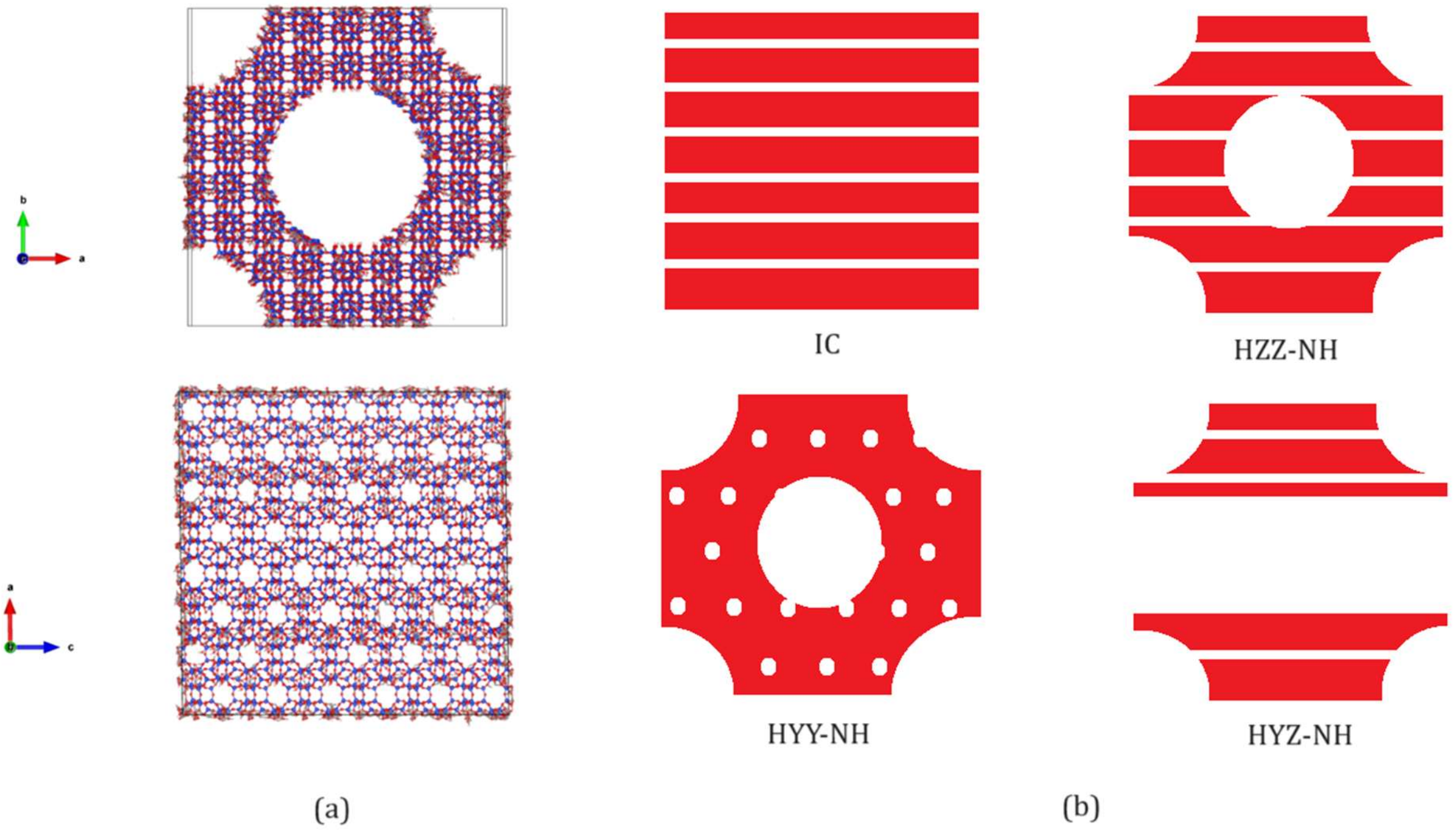

**Figure 1** (a) Simulation snapshot of the model HZZ-OH in the *a-b* (top left) and *a-c* (bottom left) planes. A mesopore at the cell center and a quarter pore at each of the corners can be identified as empty space. In the *a-c* plane, several straight channels running perpendicular to the plane of the figure can be seen as smaller empty ellipsoidal regions. (b) Schematic showing adsorbent models. Straight channels in the silicalite cell IC (top center) are shown as horizontal white lines. Sinusoidal channels are not shown for clarity. In the HZZ-NH model (top right), the mesopores run perpendicular to the straight channels, while they are parallel to the straight channels in the HYY-NH model (bottom center). In HYZ-NH, the central mesopore runs parallel to the straight channel while the corner pores are perpendicular to them.

# 3.0 RESULTS

## 3.1 Adsorption isotherms of DL hydrogen in silicalite models

In Figure 2, we plot the adsorption isotherms of hydrogen modeled using DL force-field adsorbed in different adsorbent models at 298 K. For clarity, the isotherms at lower pressures are plotted in a linear pressure scale in Figure 2(a) and the isotherms over the entire pressure range are plotted in logarithmic pressure scale in Figure 2(b). We show the amount of hydrogen adsorbed in terms of wt % of the system. The data shown here for each pressure listed in terms of number of adsorbed molecules and normalized in terms of mmol/g are included in the supplementary information. Adding mesopores in silicalite has a significant effect on the adsorption of hydrogen. All hierarchical models exhibit significantly higher amounts of hydrogen adsorbed compared to the ideal crystallite. However, the effects of the alignment of the mesopore along different directions is comparatively weaker. At lower pressures the HZZ-NH model adsorbs slightly larger amounts of hydrogen compared to HYY-NH or HYZ-NH. However, this difference is not large enough to be visible in the entire range of pressures investigated. The effect of hydroxylation results in a consistent reduction in the amounts of adsorbed hydrogen at all pressures as exhibited by the lower amounts seen for HZZ-OH model compared to that for HZZ-NH. In what follows we examine these and other differences in detail.

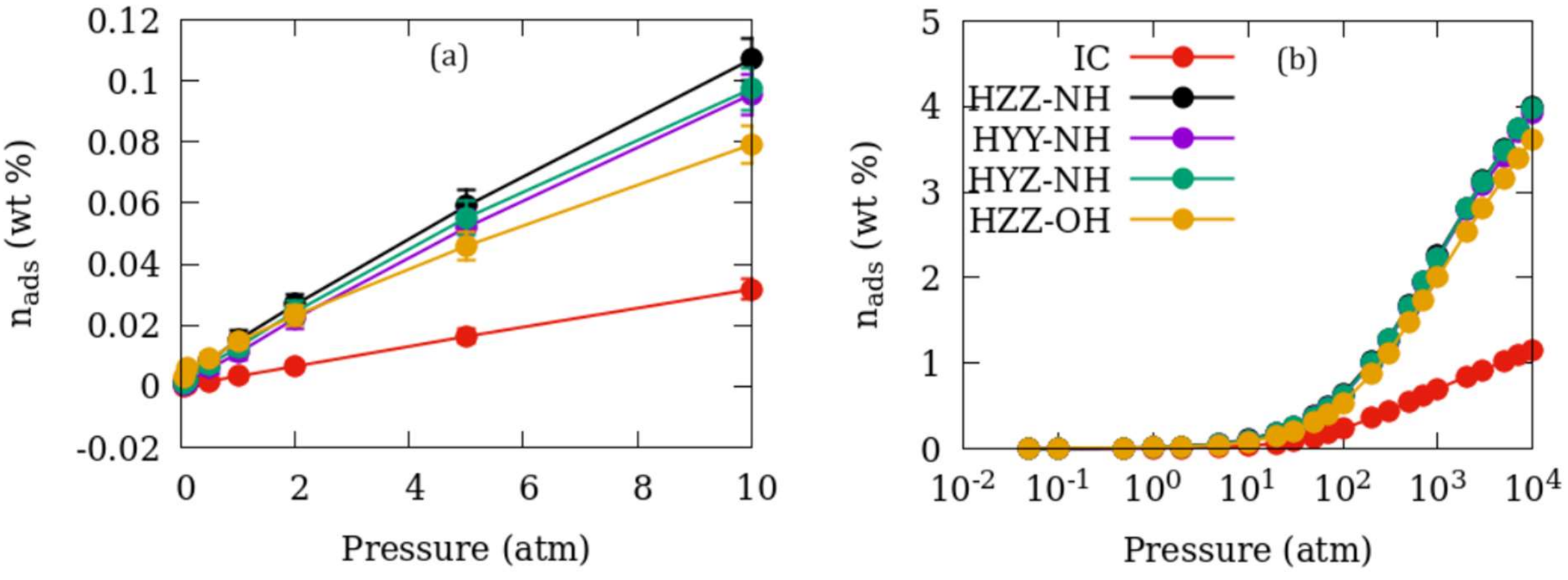

**Figure 2** Amount of hydrogen adsorbed in different models of silicalite at 298 K as indicated. Refer to Table 1 and section 2.1 for the nomenclature used. Isotherms only up to 10 atm are shown in (a) for clarity, while the complete isotherms at all pressures investigated up to 10 katm are shown in (b). Identical symbols are used to represent data in (a) and (b).

### 3.2 Gain in adsorption capacity due to mesopore inclusion

Figure 3 shows the gain in the amounts of DL hydrogen adsorbed in silicalite achieved by incorporating mesopores. The percent gain plotted here is the relative difference between the amount of hydrogen adsorbed in the hierarchical silicalite versus that in unmodified silicalite (IC) as shown in Eq. 1

$$Gain\ (\%) = \frac{n^{H}_{ads} - n^{IC}_{ads}}{n^{IC}_{ads}} \times 100 \qquad (1)$$

Here, $n^{i}_{ads}$ stands for the amount of hydrogen adsorbed in the adsorbent $i$ and H and IC respectively stand for the hierarchical and ideal crystal silicalite. The gain is particularly high at pressures below 0.5 atm. However, as the absolute amount of hydrogen adsorbed at these pressures is small, the associated uncertainty in these amounts is characteristically high. The uncertainty reduces to acceptable values at higher pressures while gain achieved remains close to 100 % for all hierarchical models of silicalite up to 10 atm. At low pressures, the hierarchical model with surface hydroxyls exhibits highest gains whereas at all pressures above 2 atm it has the smallest gain achieved in hydrogen adsorption *vis-á-vis* the unmodified silicalite. Nevertheless, the gain is always positive and above 40% at all pressures for all hierarchical models.

Compared to DL hydrogen, the gain achieved in adsorption amounts of Buch hydrogen by incorporating mesopores in silicalite is modest with up to 20% gain at low pressures progressing to about 100 % at pressures above 5 katm (see Figure 4).

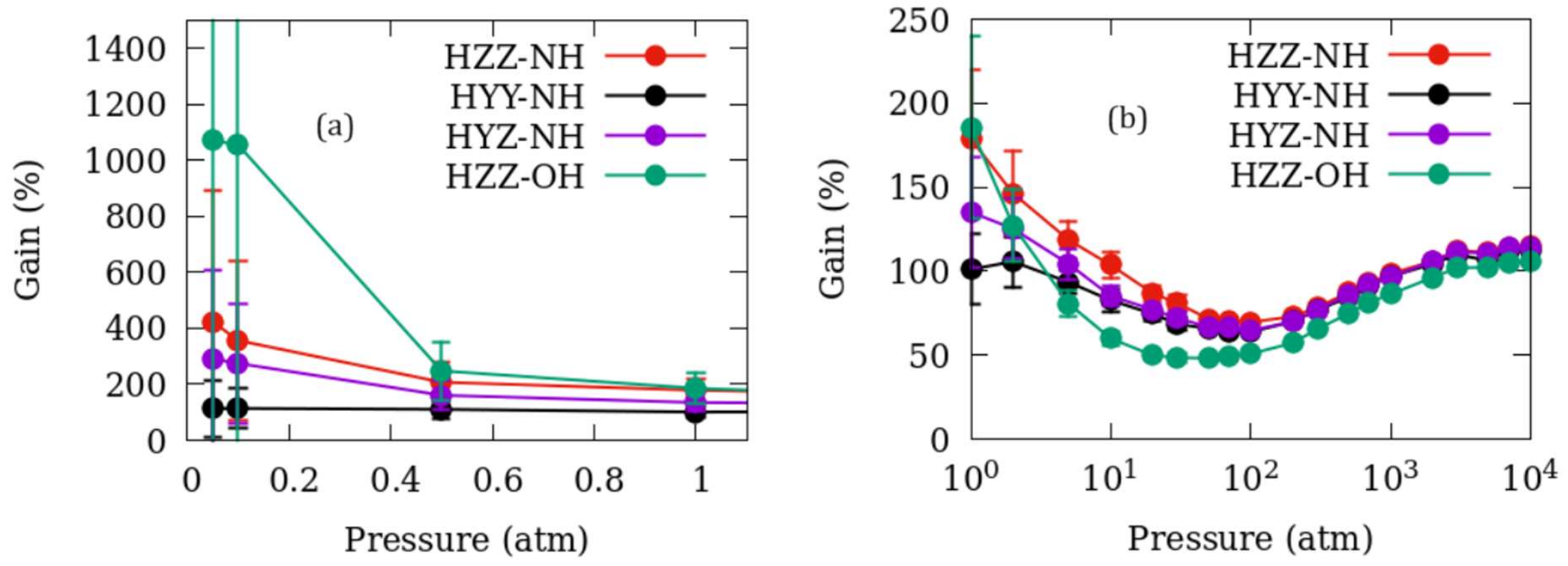

**Figure 3** Gain achieved in the amounts of hydrogen adsorbed in different models of silicalite at 298 K *vis-á-vis* unmodified silicalite (IC), when hydrogen is modeled using DL force field. Refer to Table 1 and section 2.1 for the nomenclature used. Isotherms only up to 1 atm are shown in (a) for clarity, while the complete isotherms at higher pressures are shown in (b).

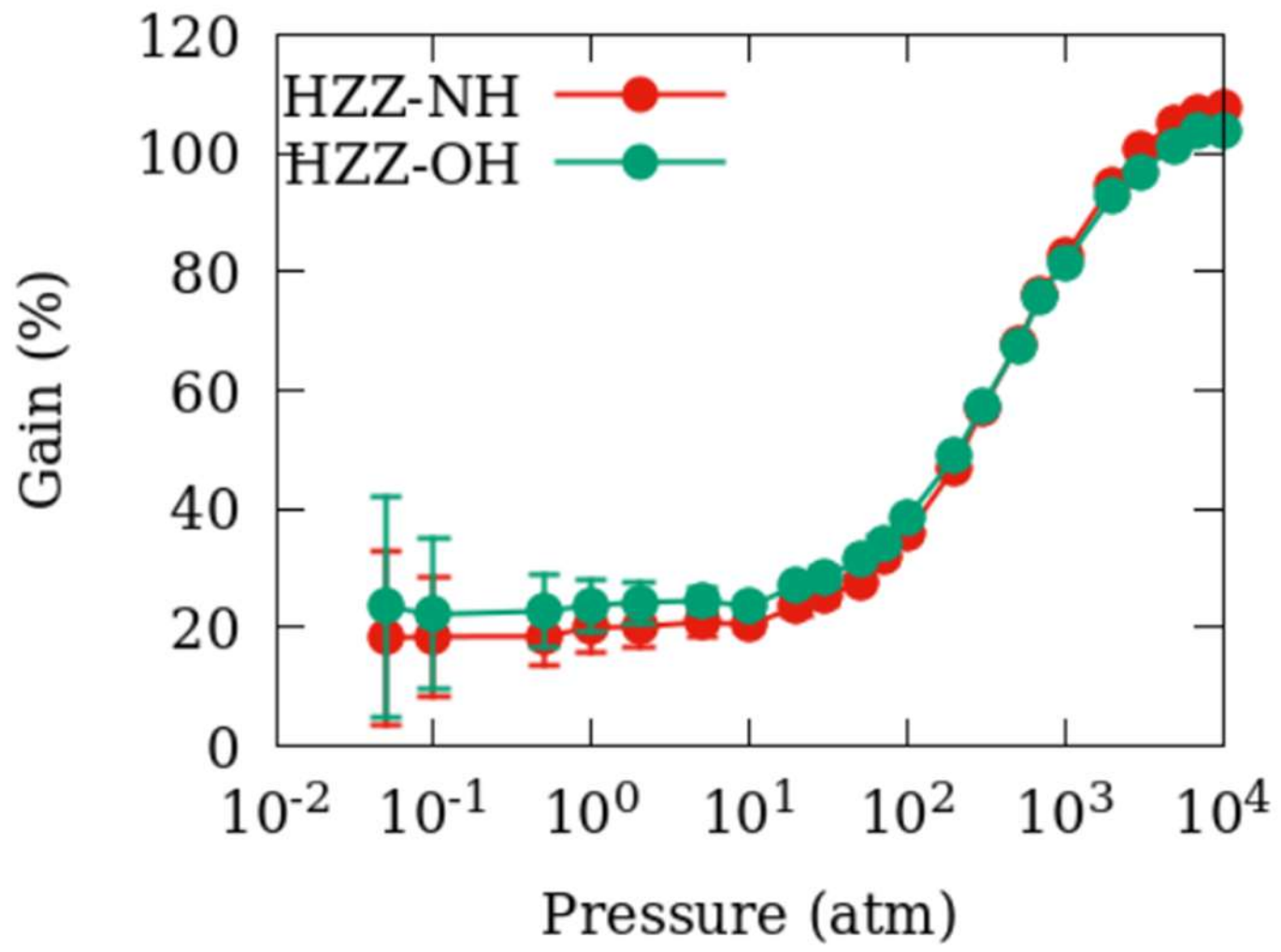

**Figure 4** Gain achieved in the amounts of hydrogen adsorbed in different models of silicalite at 298 K *vis-á-vis* unmodified silicalite, when hydrogen is modeled using Buch force field. Refer to Table 1 and section 2.1 for the nomenclature used.

## 3.3 Effects of different force-fields

In Figure 5, we show the relative difference in the amounts of hydrogen adsorbed in different models of silicalite when either the Buch or DLnc force-field is used to model hydrogen *vis-á-vis* the DL force-field. This relative difference is calculated using Eq. 2.

$$Difference\ (\%) = \frac{N_{ads}^{i} - N_{ads}^{DL}}{N_{ads}^{DL}} \times 100 \qquad (2)$$

Here, $N_{ads}^{i}$ stands for the amount of hydrogen adsorbed in a given adsorbent with the force-field $i$ (=Buch or DLnc) and $N_{ads}^{DL}$ stands for the amount of hydrogen adsorbed in the same adsorbent with DL force-field. The first four letters in the legend of Figure 5 stand for the force-field used (Buch or DLnc) while the rest stand for the adsorbent model. For $H_2$ in unmodified silicalite, the effects of using different force-fields remains below 20% at all pressures, whereas in the case of hierarchical models, these effects are much stronger particularly at low pressures. Furthermore, using the Buch force-field results in a stronger deviation with respect to the DL force field compared to just switching off the electrostatic interactions in DL (as in DLnc). This is because of the small but finite differences in the parameters used to represent the van der Waals interactions in the two force-fields.

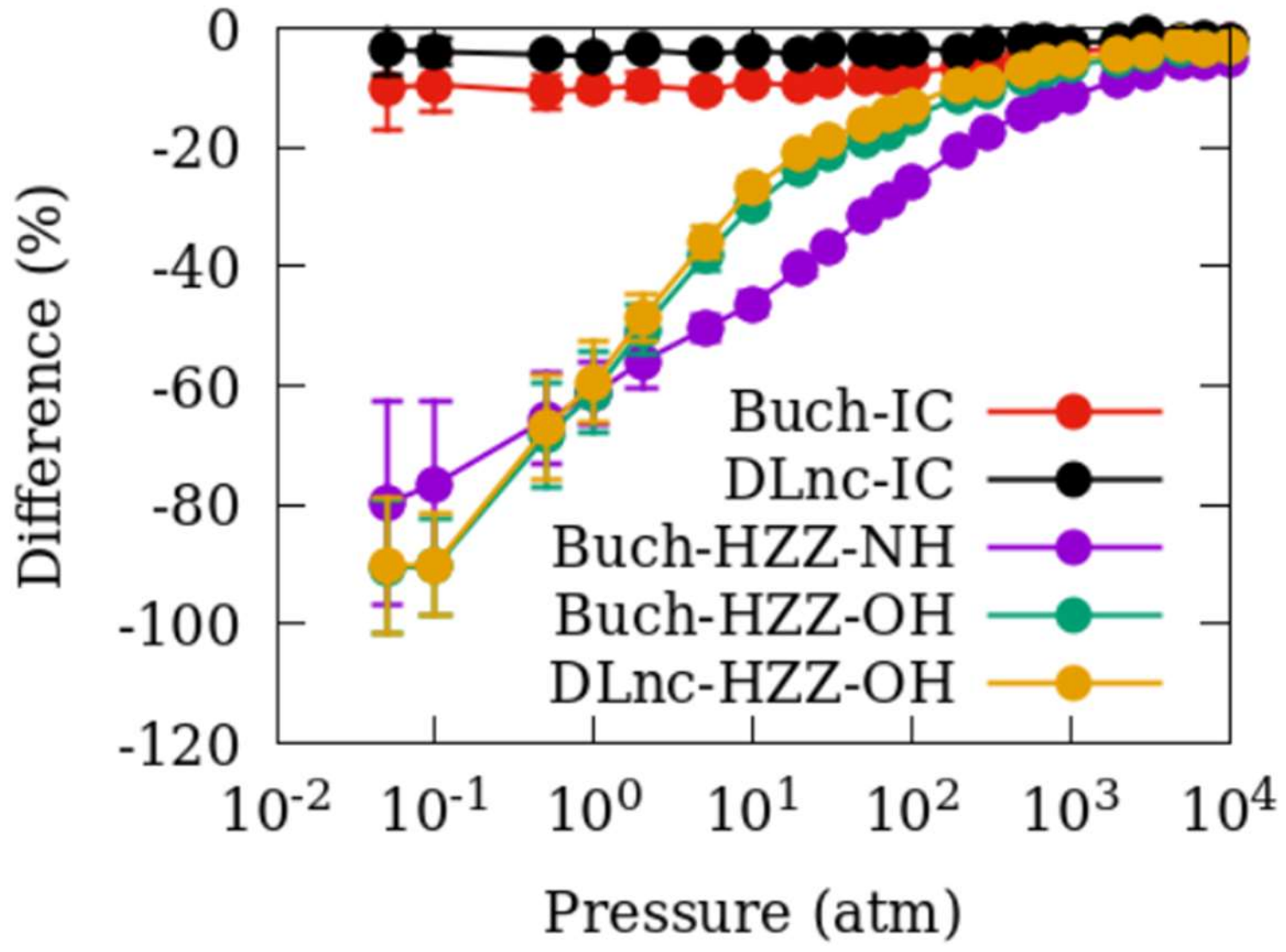

**Figure 5** Relative difference between the amounts of hydrogen adsorbed in different models of silicalite at 298 K using Buch or DLnc force-field *vis-á-vis* when DL force-field is used, calculated using Eq. 2.

## 3.4 Effects of hydroxylation

Effects of the pore surface chemistry are interrogated by comparing the relative difference (Diff-OH) in amounts of hydrogen adsorbed in hierarchical silicalite with and without pore surface terminations with hydroxyl groups (HZZ-OH and HZZ-NH) using Eq. 3.

$$Diff - OH\ (\%) = \frac{n_{ads}^{HZZ-OH} - n_{ads}^{HZZ-}}{n_{ads}^{HZZ-NH}} \times 100 \qquad (3)$$

Here, $n_{ads}^{HZZ-OH}$ and $n_{ads}^{HZZ-N}$ stand for the amount of hydrogen adsorbed in HZZ-OH and HZZ-NH using the same set of force-fields (either Buch or DL) in both cases. The relative difference is reported in Figure 6. At low pressures, hydroxylation enhances the adsorption of DL hydrogen. However, the enhancement achieved with pore surface hydroxylation has large uncertainties due to the smaller amounts of adsorption as noted earlier. At pressures above 0.5 atm, the adsorption amounts of DL hydrogen is lower in the hydroxylated pore and reaches the lowest value of -25% compared to non-hydroxylated pore at 10 atm. At pressures above this, the effects of hydroxylation decrease progressively with hydroxylated pore exhibiting lower adsorption amounts. Compared to DL hydrogen, the effects of hydroxylation on the adsorption of Buch hydrogen are insignificant. This is expected since in ClayFF, protons belonging to the hydroxyl groups that are normally closer to an adsorbed species interact with the adsorbed species only via electrostatic interactions, while the van der Waals interactions are dominated by the oxygens belonging to the hydroxyl group. This means that in absence of electrostatic interactions, as in the case of Buch hydrogen, hydroxyl groups cannot be expected to make a significant contribution to the adsorption amounts.

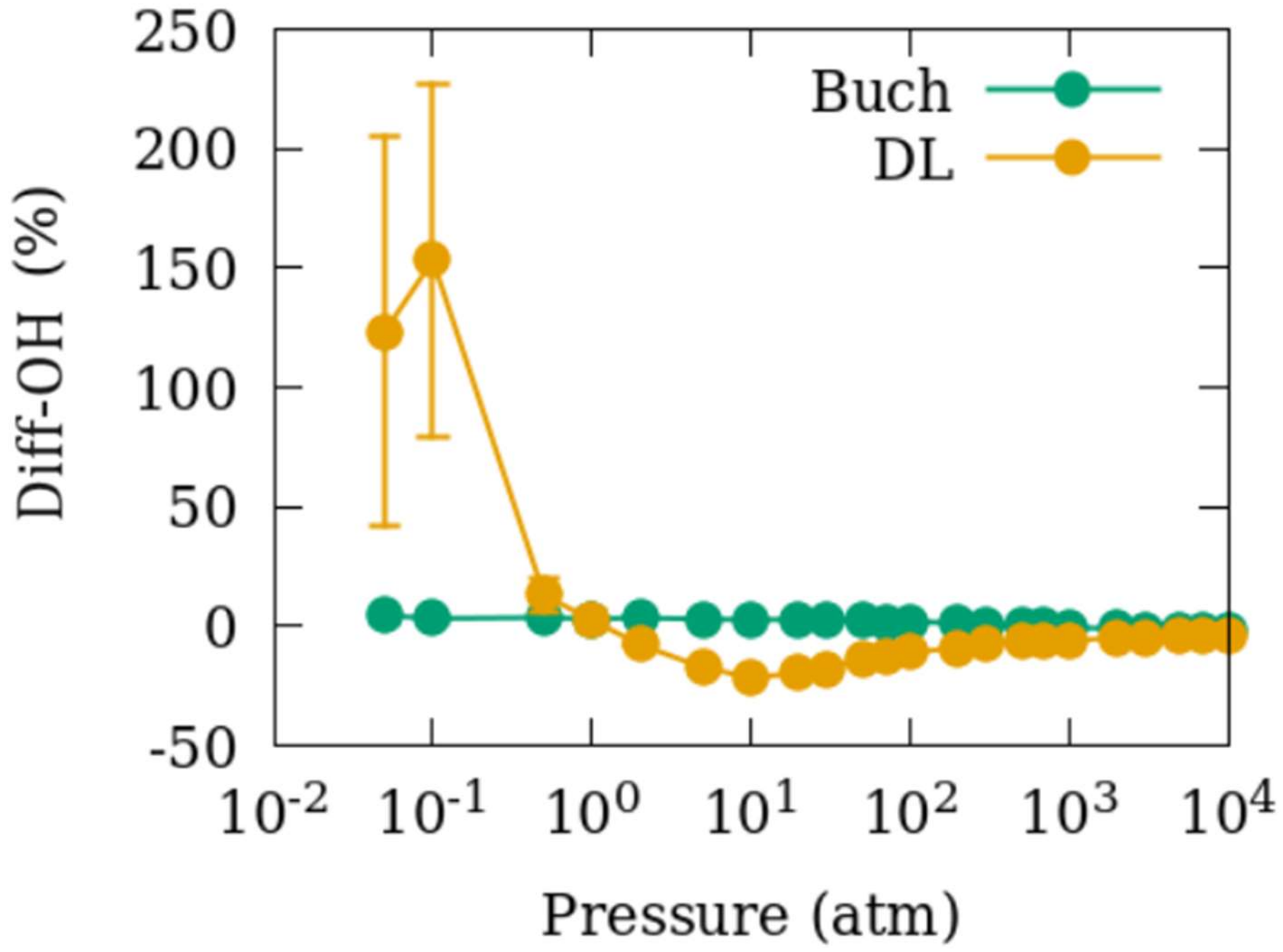

**Figure 6** Relative difference between the amounts of hydrogen adsorbed in HZZ-OH *vis-á-vis* HZZ-NH at 298 K using Buch or DL force-field. See Eq. 3.

3.5 Effects of temperature

Figure 7 shows the amount of hydrogen adsorbed in IC and HZZ-OH at different temperatures with hydrogen modeled using DL or Buch force fields. For clarity, we have included data corresponding to the pressures of 1 and 10 atm in the plots. The data corresponding to two other pressures for which temperature dependence is investigated can be found in the supplementary file. As expected, the adsorption amounts decrease as the temperature is increased. While the temperature dependence of hydrogen in the IC model is similar for both the force-fields, for HZZ-OH model, DL force-field exhibits a slightly stronger temperature dependence.

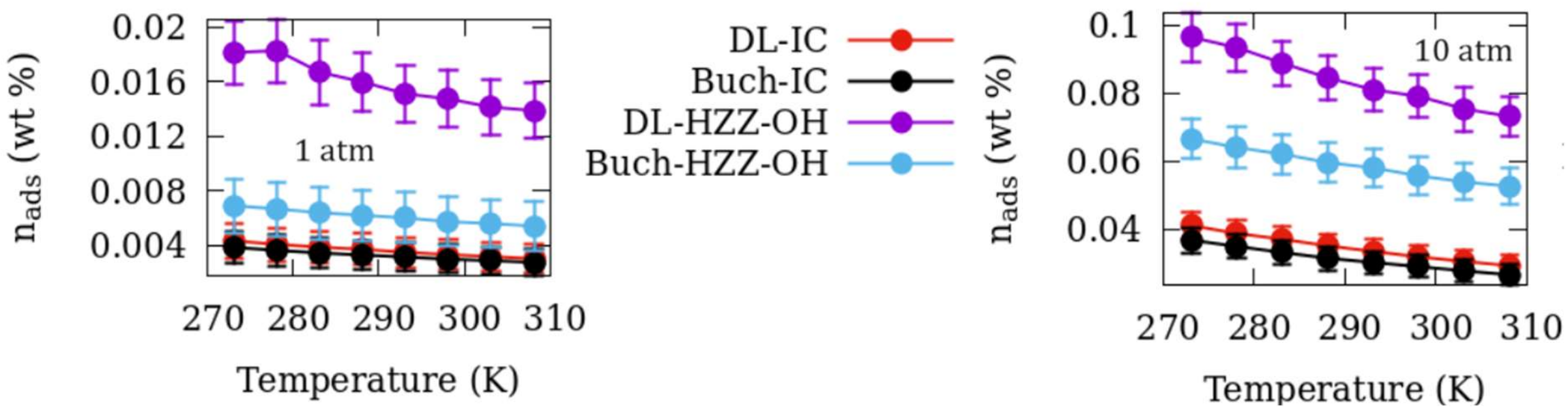

**Figure 7** Amounts of hydrogen adsorbed at a pressure of 1 atm (left) and 10 atm (right) as a function of temperature.

## 4.0 DISCUSSION

Fujiwara et al. studied hydrogen adsorption in ZSM-5 zeolite grafted with 1,4-bis(hydroxydimethylsilyl)benzene to seal the micropores [42]. At 77 K and 0.98 atm; they obtained a hydrogen adsorption of 6.04 mg/g (~0.6 wt%) in the unsealed ZSM-5, while the sealed ZSM-5 exhibited a much lower adsorption. More recently, Radola et al obtained similar values for adsorption of hydrogen in MFI type zeolite under the same conditions of temperature and pressure [43]. In this work we have used an all-silica form of ZSM-5 (MFI framework type), which is structurally identical to the adsorbent used in ref. 42 and 43. To our knowledge, no experimental or computational work has reported hydrogen adsorption in these materials at 298 K. In our work, the adsorption amount of 0.6 wt% is achieved at about 1 katm of pressure of hydrogen. This is a significantly low amount of adsorbed hydrogen. However, as our results demonstrate, the adsorption amounts can be significantly improved by incorporating a hierarchical pore structure.

The role of electrostatics and hydroxylation of pore surface is elucidated via comparative simulations employing different force-fields. The effects of these attributes are particularly strong at lower pressures. In the simulations, the interaction of hydrogen of the hydroxyl groups with any other species is exclusively accounted for by Coulombic interactions. Therefore, it can be expected that the effects of electrostatic interactions and the surface hydroxylation are inter-related. In nanoporous materials with different type of pore spaces, it can be insightful to study the ratio of molecules adsorbed in different pore types [31, 44]. In this work the hierarchical silicalite adsorbents have mainly two types of pores – micropores (0.55 nm wide) and mesopores (4 nm wide). In Figure 8, we plot the fraction of all hydrogen molecules that are adsorbed in the mesopores in the hierarchical models HZZ-NH and HZZ-OH, for the case of DL force-field. In case of non-hydroxylated model (HZZ-NH) the adsorption is predominantly in the mesopores (ratio >0.5) whereas in the hydroxylated model (HZZ-OH) it is mostly in micropores (ratio<0.5) up to 10 atm. This means that in the hydroxylated model at low pressures hydrogen is preferentially adsorbed in the micropores. This preference for the micropores in the hydroxylated model is because of the -OH groups present on the surface of mesopores.

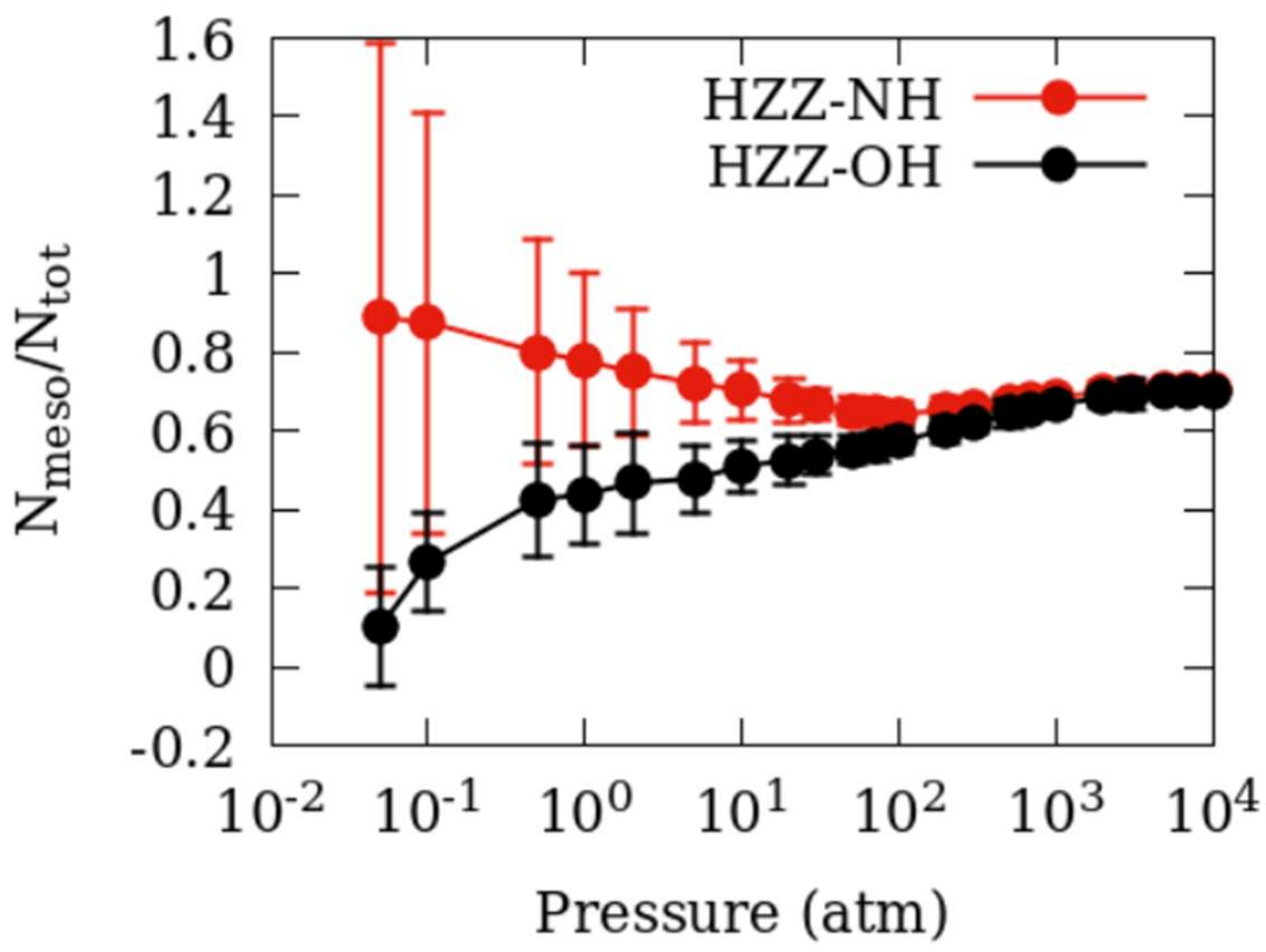

**Figure 8** Fraction of hydrogen molecules adsorbed in the mesopores in the simulations employing DL force field to model hydrogen. A value of greater than 0.5 indicates hydrogen molecules are predominantly adsorbed in mesopores.

While the effects reported here are observed for silicalite, we believe they should also be applicable to a wide variety of nanoporous materials. Hydrogen adsorption capacity of nanomaterials can be enhanced by a proper control of pore structure and surface hydroxylation. This should help in the selection and optimization of nanoporous materials for hydrogen storage.

## 5.0 CONCLUSIONS

In this work we report a grand canonical Monte Carlo (GCMC) simulation study on the adsorption of hydrogen in hierarchical models of silicalite that incorporates 4 nm wide mesopores in addition to the 0.5 nm wide micropores present in the unmodified silicalite at 298 K. Using different force fields to model hydrogen that differ in the use of electrostatic interactions to account for the molecular quadrupole moment, we documented the effects of these interactions on the adsorption amounts. Finally, the mesopores in the hierarchical models of silicalite exhibit different surface chemistries in terms of presence or absence of hydroxyl groups, which helps us determine the role played by the pore surface chemistry in hydrogen adsorption. Our results suggest that incorporating mesopores in silicalite can enhance the hydrogen adsorption significantly. The gain in adsorption amounts is at least 20% if electrostatic interactions are not included in the simulations and up to 100 % otherwise. Incorporating the quadrupolar moment of hydrogen in the simulations by accounting for the electrostatic interactions results in higher amounts of hydrogen adsorption by close to 100% at lower pressures for hierarchical silicalite whereas for unmodified silicalite, the effect of electrostatic interactions less significant at all pressures. Hydroxylating the mesopore surface in hierarchical silicalite results in an enhancement in hydrogen adsorption at pressures below 1 atm, whereas at higher pressures, suppression by up to 20 % is observed. When the mesopore surface is hydroxylated, hydrogen is predominantly adsorbed in the micropores at lower pressures whereas in absence of hydroxyl groups, the adsorption is predominantly in the mesopores at all pressures. We also studied the temperature dependence of adsorption at selected pressures that revealed the expected decrease in adsorption amounts at higher temperatures. Results reported here can be useful in the engineering, selection, and optimization of nanoporous materials for hydrogen storage.

**ACKNOWLEDGEMENTS**

Funding for SG and DRC was provided by the State of Ohio through the Third Frontier Ohio Research Scholar Program. We would like to acknowledge STFC's Daresbury Laboratory for providing the packages DL-Monte and DL-Poly, which were used in this work. The figures in this manuscript were made using the freely available visualization and plotting softwares VESTA [45] and Gnuplot version 5.2 [46].

**CONFLICT OF INTEREST**

The authors declare that there is no conflict of interest.

## SUPPLEMENTARY INFORMATION

This supplement contains the following information regarding the simulations reported in the main article.

1. Force-field parameters used in the simulations,
2. Complete adsorption isotherms of hydrogen in all adsorbent models encompassing all the simulations listed in Table 1 in the main article.
3. Temperature variation of adsorption amounts of hydrogen at four representative pressures in IC and HZZ-OH models.

1. Force-Field parameters

As stated in the main article this work used ClayFF [SR1] force-field to represent the adsorbent models and three different force-fields to represent the hydrogen molecules. The simulations used the following expression for the interaction energy.

$$U_{ij} = 4\varepsilon_{ij}\left[\left(\frac{\sigma_{ij}}{r_{ij}}\right)^{12} - \left(\frac{\sigma_{ij}}{r_{ij}}\right)^{6}\right] + \frac{q_i q_j}{4\pi\varepsilon_0 r_{ij}} \qquad \text{(S1)}$$

where $\varepsilon_{ij}$ is the depth of the potential well, $\sigma_{ij}$ is the distance at which the intermolecular potential between the atoms $i$ and $j$ becomes zero, the van der Waals radius, and $r_{ij}$ is the distance between atoms $i$, and $j$, $q_i$, and $q_j$ are the charges of the $i$ and $j$ atoms. The parameters $\varepsilon_{ii}$ and $\sigma_{ii}$ and $q_i$ for the like pairs ($i$=$j$) of different atoms and different force-fields are provided in Tables S1 and S2. The parameters for unlike pairs ($i$≠$j$) are obtained using the following mixing rules.

$$\varepsilon_{ij} = \sqrt{\varepsilon_{ii} \times \varepsilon_{jj}} \qquad \text{(S2)}$$

$$\sigma_{ij} = \frac{\sigma_{ii}+\sigma_{jj}}{2} \qquad \text{(S3)}$$

### 1.1 Models of Hydrogen molecule

As mentioned in the main article, hydrogen molecules are modeled in two different formalisms – Buch [SR2} and Darkrim-Levesque (DL) [SR3] corresponding to two force-fields with these names. Further, putting off the charges in DL model resulted in a third set of force-field parameters DLnc. The two models of hydrogen are shown in Figure S1.

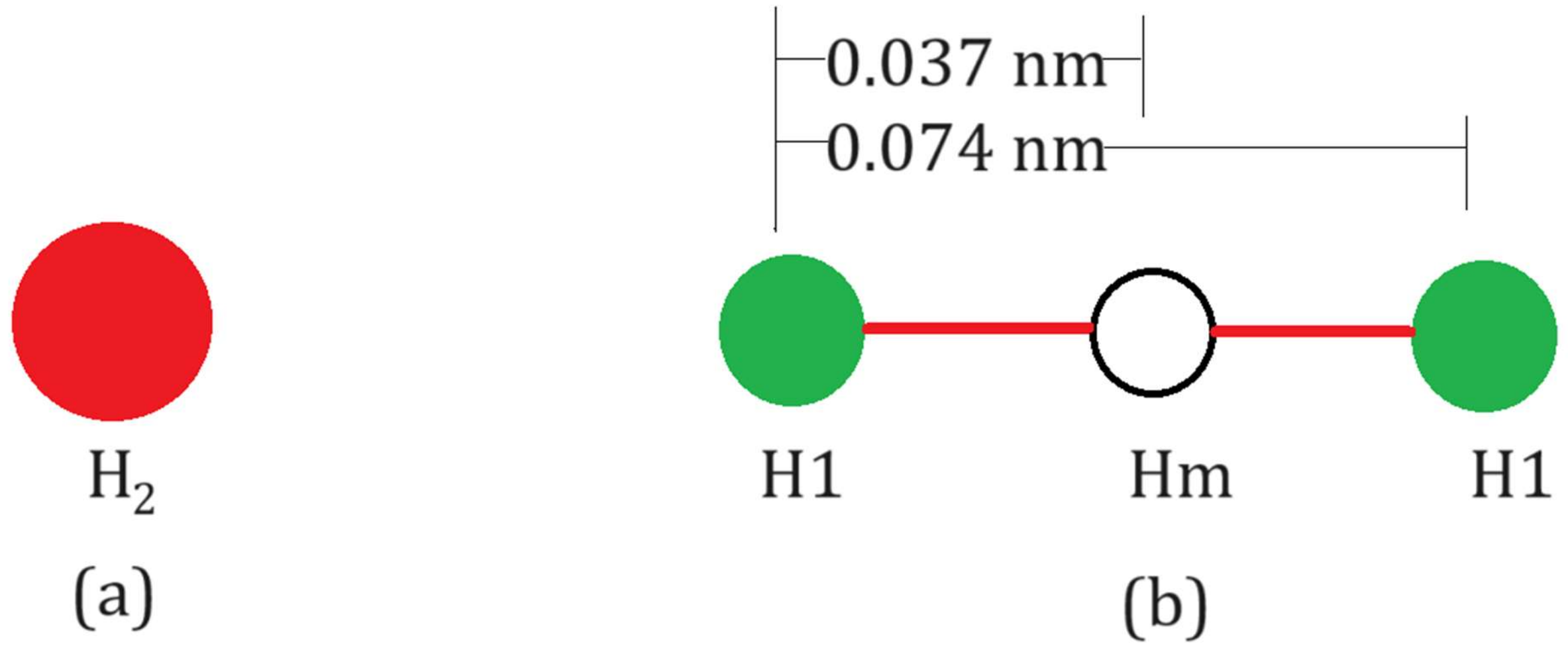

Figure S1. Models of hydrogen molecule in (a) Buch and (b) Darkrim-Levesque (DL) formalism

Table S1. Force-field parameters of adsorbent atoms in ClayFF formalism

| Atom | $\varepsilon_{ii}$ (kJ/mol) | $\sigma_{ii}$ (Å) | $q_i$ ($q/e$) |
|---|---|---|---|
| Si | 0.000008 | 3.301 | +2.10 |
| O | 0.650198 | 3.166 | -1.05 |
| Oh[a] | 0.650198 | 3.166 | -0.950 |
| H[b] | - | - | +0.425 |

[a] Atom Oh is the oxygen belonging to the hydroxyl group that it forms with the H atom.

[b] In ClayFF force-field, the hydrogen atom belonging to the hydroxyl group interacts with other atoms only via Coulombic interactions and has no van der Waals contributions.

Table S2. Force-field parameters of hydrogen molecules in different force-fields

| Force-field[c] | Atom/pseudo-atom | $\varepsilon_{ii}$ (kJ/mol) | $\sigma_{ii}$ (Å) | $q_i$ ($q/e$) |
|---|---|---|---|---|
| Buch | $H_2$ | 0.28435 | 2.96 | 0 |
| DL[d] | Hm | 0.30514 | 2.958 | -0.936 |
| | H1 | - | - | +0.468 |
| DLnc | Hm | 0.30514 | 2.958 | 0 |
| | H1 | - | - | 0 |

[c] See Figure S1 for a description of the location of different atoms constituting the hydrogen molecule.

[b] In DL (and DLnc) force-field, the atom H1 interacts with other atoms only via Coulombic interactions and has no van der Waals contributions.

2. Adsorption Isotherms

The complete adsorption isotherm data for all the simulations listed in Table 1 in the main article are listed in the following tables in different units as indicated. In these tables, the first column shows the pressure in katm, the second and third columns show respectively the average number of molecules adsorbed in the adsorbent at the given pressure and the corresponding standard deviation. The fourth and fifth columns show respectively, the adsorbed amounts of hydrogen expressed in mmol/g and the corresponding standard deviation. The last two (sixth and seventh) columns show the adsorbed amounts of hydrogen as wt%.

Table S3. Adsorption isotherm of Buch hydrogen in silicalite model IC at 298 K.

| Pressure (katm) | N (molecules) | dN (molecules) | $n_{ads}$(mmol/g) | $dn_{ads}$(mmol/g) | $n_{ads}$ (wt%) | $dn_{ads}$(wt%) |
|---|---|---|---|---|---|---|
| 5.00E-05 | 0.414 | 0.6409 | 7.49E-04 | 1.16E-03 | 1.51E-04 | 2.34E-04 |
| 1.00E-04 | 0.8335 | 0.9089 | 1.51E-03 | 1.64E-03 | 3.04E-04 | 3.31E-04 |
| 5.00E-04 | 4.1113 | 2.0498 | 7.44E-03 | 3.71E-03 | 1.50E-03 | 7.47E-04 |
| 1.00E-03 | 8.159 | 2.8666 | 1.48E-02 | 5.19E-03 | 2.97E-03 | 1.05E-03 |
| 2.00E-03 | 16.2671 | 4.1301 | 2.94E-02 | 7.47E-03 | 5.93E-03 | 1.51E-03 |
| 5.00E-03 | 40.1177 | 6.3312 | 7.26E-02 | 1.15E-02 | 1.46E-02 | 2.31E-03 |
| 1.00E-02 | 79.27 | 8.8599 | 0.143406 | 1.60E-02 | 2.89E-02 | 3.23E-03 |
| 2.00E-02 | 151.0671 | 12.3726 | 0.273292 | 2.24E-02 | 5.50E-02 | 4.51E-03 |
| 3.00E-02 | 218.1843 | 15.1195 | 0.394713 | 2.74E-02 | 7.95E-02 | 5.50E-03 |
| 5.00E-02 | 340.1218 | 17.0779 | 0.615307 | 3.09E-02 | 0.123849 | 6.21E-03 |
| 7.00E-02 | 443.4952 | 21.361 | 0.802318 | 3.86E-02 | 0.16143 | 7.76E-03 |
| 0.1 | 582.2325 | 22.5306 | 1.053304 | 4.08E-02 | 0.211823 | 8.18E-03 |
| 0.2 | 916.782 | 27.6013 | 1.65853 | 4.99E-02 | 0.33313 | 1.00E-02 |
| 0.3 | 1131.84 | 25.5952 | 2.047588 | 4.63E-02 | 0.410954 | 9.26E-03 |
| 0.5 | 1441.242 | 25.5717 | 2.60732 | 4.63E-02 | 0.522706 | 9.23E-03 |
| 0.7 | 1637.242 | 22.6164 | 2.9619 | 4.09E-02 | 0.593369 | 8.15E-03 |
| 1 | 1852.991 | 23.3951 | 3.352207 | 4.23E-02 | 0.671036 | 8.42E-03 |
| 2 | 2252.427 | 25.2902 | 4.074818 | 4.58E-02 | 0.814509 | 9.07E-03 |
| 3 | 2472.613 | 24.0114 | 4.47315 | 4.34E-02 | 0.89342 | 8.60E-03 |
| 5 | 2755.856 | 20.9586 | 4.98556 | 3.79E-02 | 0.994745 | 7.49E-03 |
| 7 | 2938.405 | 17.762 | 5.315805 | 3.21E-02 | 1.059939 | 6.34E-03 |
| 10 | 3137.892 | 17.134 | 5.676693 | 3.10E-02 | 1.131083 | 6.11E-03 |

Table S4. Adsorption isotherm of DL hydrogen in silicalite model IC at 298 K.

| Pressure (katm) | N (molecules) | dN (molecules) | $n_{ads}$(mmol/g) | $dn_{ads}$(mmol/g) | $n_{ads}$ (wt%) | $dn_{ads}$(wt%) |
|---|---|---|---|---|---|---|
| 5.00E-05 | 0.4599 | 0.6757 | 8.32E-04 | 1.22E-03 | 1.68E-04 | 2.46E-04 |
| 1.00E-04 | 0.9205 | 0.9582 | 1.67E-03 | 1.73E-03 | 3.36E-04 | 3.49E-04 |
| 5.00E-04 | 4.6044 | 2.1578 | 8.33E-03 | 3.90E-03 | 1.68E-03 | 7.87E-04 |
| 1.00E-03 | 9.0889 | 3.0364 | 1.64E-02 | 5.49E-03 | 3.31E-03 | 1.11E-03 |
| 2.00E-03 | 18.0087 | 4.1694 | 3.26E-02 | 7.54E-03 | 6.57E-03 | 1.52E-03 |
| 5.00E-03 | 44.7647 | 6.3682 | 8.10E-02 | 1.15E-02 | 1.63E-02 | 2.32E-03 |
| 1.00E-02 | 87.1641 | 9.1709 | 0.157687 | 1.66E-02 | 3.18E-02 | 3.34E-03 |
| 2.00E-02 | 167.3838 | 12.5626 | 0.30281 | 2.27E-02 | 6.10E-02 | 4.57E-03 |
| 3.00E-02 | 239.4142 | 15.6302 | 0.433119 | 2.83E-02 | 8.72E-02 | 5.69E-03 |
| 5.00E-02 | 371.0343 | 18.7265 | 0.67123 | 3.39E-02 | 0.13509 | 6.81E-03 |
| 7.00E-02 | 483.0761 | 21.3501 | 0.873923 | 3.86E-02 | 0.175812 | 7.76E-03 |
| 0.1 | 628.344 | 22.2392 | 1.136724 | 4.02E-02 | 0.22856 | 8.07E-03 |
| 0.2 | 982.2926 | 23.2961 | 1.777044 | 4.21E-02 | 0.35685 | 8.43E-03 |
| 0.3 | 1204.376 | 22.4559 | 2.178812 | 4.06E-02 | 0.437176 | 8.12E-03 |
| 0.5 | 1509.979 | 27.3524 | 2.731671 | 4.95E-02 | 0.547499 | 9.86E-03 |
| 0.7 | 1708.263 | 25.5623 | 3.090382 | 4.62E-02 | 0.618949 | 9.20E-03 |
| 1 | 1928.552 | 17.5497 | 3.488903 | 3.17E-02 | 0.698209 | 6.31E-03 |
| 2 | 2331.774 | 22.8652 | 4.218361 | 4.14E-02 | 0.84296 | 8.20E-03 |
| 3 | 2527.455 | 24.4132 | 4.572363 | 4.42E-02 | 0.913054 | 8.74E-03 |
| 5 | 2845.022 | 20.5532 | 5.146867 | 3.72E-02 | 1.026599 | 7.34E-03 |
| 7 | 3014.617 | 17.002 | 5.453678 | 3.08E-02 | 1.087131 | 6.06E-03 |
| 10 | 3210.685 | 21.0075 | 5.808382 | 3.80E-02 | 1.157019 | 7.48E-03 |

Table S5. Adsorption isotherm of DLnc hydrogen in silicalite model IC at 298 K.

| Pressure (katm) | N (molecules) | dN (molecules) | $n_{ads}$(mmol/g) | $dn_{ads}$(mmol/g) | $n_{ads}$ (wt%) | $dn_{ads}$(wt%) |
|---|---|---|---|---|---|---|
| 5.00E-05 | 0.443 | 0.6707 | 8.01E-04 | 1.21E-03 | 1.62E-04 | 2.45E-04 |
| 1.00E-04 | 0.884 | 0.9397 | 1.60E-03 | 1.70E-03 | 3.22E-04 | 3.43E-04 |
| 5.00E-04 | 4.399 | 2.1057 | 7.96E-03 | 3.81E-03 | 1.60E-03 | 7.68E-04 |
| 1.00E-03 | 8.6597 | 2.9329 | 1.57E-02 | 5.31E-03 | 3.16E-03 | 1.07E-03 |
| 2.00E-03 | 17.3503 | 4.0612 | 3.14E-02 | 7.35E-03 | 6.33E-03 | 1.48E-03 |
| 5.00E-03 | 42.7859 | 6.4364 | 7.74E-02 | 1.16E-02 | 1.56E-02 | 2.35E-03 |
| 1.00E-02 | 83.8956 | 9.0713 | 0.151774 | 1.64E-02 | 3.06E-02 | 3.31E-03 |
| 2.00E-02 | 160.0277 | 12.1236 | 0.289503 | 2.19E-02 | 5.83E-02 | 4.41E-03 |
| 3.00E-02 | 231.2495 | 14.4275 | 0.418348 | 2.61E-02 | 8.42E-02 | 5.25E-03 |
| 5.00E-02 | 357.8723 | 16.9039 | 0.647419 | 3.06E-02 | 0.13030 | 6.15E-03 |
| 7.00E-02 | 463.9291 | 19.5897 | 0.839284 | 3.54E-02 | 0.16885 | 7.12E-03 |
| 0.1 | 607.3493 | 22.8371 | 1.098742 | 4.13E-02 | 0.22094 | 8.29E-03 |
| 0.2 | 945.0783 | 24.9304 | 1.709721 | 4.51E-02 | 0.34337 | 9.03E-03 |
| 0.3 | 1174.36 | 22.0792 | 2.124509 | 3.99E-02 | 0.42632 | 7.98E-03 |
| 0.5 | 1474.62 | 24.4619 | 2.667703 | 4.43E-02 | 0.53474 | 8.82E-03 |
| 0.7 | 1668.198 | 22.4651 | 3.017901 | 4.06E-02 | 0.60452 | 8.09E-03 |
| 1 | 1882.044 | 37.3477 | 3.404765 | 6.76E-02 | 0.68148 | 1.34E-02 |
| 2 | 2279.858 | 22.0914 | 4.124442 | 4.00E-02 | 0.82434 | 7.92E-03 |
| 3 | 2507.312 | 24.3529 | 4.535923 | 4.41E-02 | 0.90584 | 8.72E-03 |
| 5 | 2778.154 | 25.2282 | 5.025899 | 4.56E-02 | 1.00271 | 9.01E-03 |
| 7 | 2968.49 | 19.7275 | 5.370232 | 3.57E-02 | 1.07067 | 7.04E-03 |
| 10 | 3145.253 | 25.6402 | 5.69001 | 4.64E-02 | 1.13370 | 9.14E-03 |

Table S6. Adsorption isotherm of Buch hydrogen in hierarchical silicalite HZZ-NH at 298 K.

| Pressure (katm) | N (molecules) | dN (molecules) | $n_{ads}$(mmol/g) | $dn_{ads}$(mmol/g) | $n_{ads}$ (wt%) | $dn_{ads}$(wt%) |
|---|---|---|---|---|---|---|
| 5.00E-05 | 0.4895 | 0.6977 | 1.47E-03 | 2.09E-03 | 2.96E-04 | 4.21E-04 |
| 1.00E-04 | 0.9871 | 0.9917 | 2.96E-03 | 2.97E-03 | 5.96E-04 | 5.99E-04 |
| 5.00E-04 | 4.8713 | 2.2187 | 1.46E-02 | 6.65E-03 | 2.94E-03 | 1.34E-03 |
| 1.00E-03 | 9.7893 | 3.1137 | 2.93E-02 | 9.33E-03 | 5.91E-03 | 1.88E-03 |
| 2.00E-03 | 19.54 | 4.4118 | 5.85E-02 | 1.32E-02 | 1.18E-02 | 2.66E-03 |
| 5.00E-03 | 48.4904 | 6.7502 | 0.145259 | 2.02E-02 | 2.93E-02 | 4.07E-03 |
| 1.00E-02 | 95.4681 | 9.6807 | 0.285987 | 2.90E-02 | 5.76E-02 | 5.84E-03 |
| 2.00E-02 | 186.5603 | 13.1842 | 0.558865 | 3.95E-02 | 0.112501 | 7.94E-03 |
| 3.00E-02 | 273.2574 | 15.414 | 0.818578 | 4.62E-02 | 0.164696 | 9.27E-03 |
| 5.00E-02 | 434.4332 | 21.5573 | 1.3014 | 6.46E-02 | 0.261585 | 1.29E-02 |
| 7.00E-02 | 585.0022 | 22.5231 | 1.752449 | 6.75E-02 | 0.351928 | 1.35E-02 |
| 0.1 | 790.861 | 28.0326 | 2.369126 | 8.40E-02 | 0.475181 | 1.68E-02 |
| 0.2 | 1348.136 | 31.8258 | 4.038514 | 9.53E-02 | 0.807311 | 1.89E-02 |
| 0.3 | 1777.775 | 37.6691 | 5.325554 | 0.112843 | 1.061862 | 2.23E-02 |
| 0.5 | 2418.995 | 36.3419 | 7.246411 | 0.108867 | 1.439348 | 2.13E-02 |
| 0.7 | 2882.794 | 38.8586 | 8.635779 | 0.116406 | 1.710597 | 2.27E-02 |
| 1 | 3387.282 | 38.8584 | 10.14704 | 0.116405 | 2.003952 | 2.25E-02 |
| 2 | 4387.34 | 35.0495 | 13.14284 | 0.104995 | 2.58033 | 2.01E-02 |
| 3 | 4968.062 | 37.4324 | 14.88247 | 0.112134 | 2.911925 | 2.13E-02 |
| 5 | 5656.441 | 38.3525 | 16.9446 | 0.11489 | 3.302081 | 2.16E-02 |
| 7 | 6084.619 | 28.6521 | 18.22726 | 8.58E-02 | 3.543183 | 1.61E-02 |
| 10 | 6519.286 | 31.0766 | 19.52936 | 9.31E-02 | 3.786713 | 1.74E-02 |

Table S7. Adsorption isotherm of DL hydrogen in hierarchical silicalite HZZ-NH at 298 K.

| Pressure (katm) | N (molecules) | dN (molecules) | $n_{ads}$(mmol/g) | $dn_{ads}$(mmol/g) | $n_{ads}$ (wt%) | $dn_{ads}$(wt%) |
|---|---|---|---|---|---|---|
| 5.00E-05 | 2.4101 | 1.3866 | 7.22E-03 | 4.15E-03 | 1.45E-03 | 8.37E-04 |
| 1.00E-04 | 4.2013 | 1.7474 | 1.26E-02 | 5.23E-03 | 2.54E-03 | 1.05E-03 |
| 5.00E-04 | 14.1622 | 3.2995 | 4.24E-02 | 9.88E-03 | 8.55E-03 | 1.99E-03 |
| 1.00E-03 | 25.3356 | 4.7281 | 7.59E-02 | 1.42E-02 | 1.53E-02 | 2.85E-03 |
| 2.00E-03 | 44.3553 | 5.6727 | 0.132872 | 1.70E-02 | 2.68E-02 | 3.42E-03 |
| 5.00E-03 | 97.8369 | 9.115 | 0.293083 | 2.73E-02 | 5.90E-02 | 5.50E-03 |
| 1.00E-02 | 177.4606 | 11.5751 | 0.531606 | 3.47E-02 | 0.10702 | 6.97E-03 |
| 2.00E-02 | 312.6113 | 15.6337 | 0.936467 | 4.68E-02 | 0.188371 | 9.40E-03 |
| 3.00E-02 | 433.0857 | 17.6859 | 1.297364 | 5.30E-02 | 0.260776 | 1.06E-02 |
| 5.00E-02 | 636.2363 | 23.6498 | 1.905928 | 7.08E-02 | 0.382632 | 1.42E-02 |
| 7.00E-02 | 822.5887 | 26.9336 | 2.46417 | 8.07E-02 | 0.49415 | 1.61E-02 |
| 0.1 | 1065.723 | 29.3899 | 3.19251 | 8.80E-02 | 0.639273 | 1.75E-02 |
| 0.2 | 1698.218 | 36.8482 | 5.08723 | 0.110384 | 1.014825 | 2.18E-02 |
| 0.3 | 2150.317 | 39.9996 | 6.441552 | 0.119824 | 1.281529 | 2.35E-02 |
| 0.5 | 2825.821 | 34.34 | 8.46511 | 0.10287 | 1.677357 | 2.00E-02 |
| 0.7 | 3299.642 | 39.6304 | 9.884501 | 0.118718 | 1.953115 | 2.30E-02 |
| 1 | 3826.303 | 33.7454 | 11.46218 | 0.101089 | 2.257817 | 1.95E-02 |
| 2 | 4801.424 | 35.4053 | 14.38328 | 0.106061 | 2.817005 | 2.02E-02 |
| 3 | 5368.88 | 33.3268 | 16.08317 | 9.98E-02 | 3.139481 | 1.89E-02 |
| 5 | 6001.188 | 38.927 | 17.97733 | 0.116611 | 3.496298 | 2.19E-02 |
| 7 | 6444.321 | 34.9655 | 19.30479 | 0.104744 | 3.7448 | 1.96E-02 |
| 10 | 6895.585 | 28.578 | 20.65661 | 8.56E-02 | 3.99655 | 1.59E-02 |

Table S8. Adsorption isotherm of DL hydrogen in hierarchical silicalite HYY-NH at 298 K.

| Pressure (katm) | N (molecules) | dN (molecules) | $n_{ads}$(mmol/g) | $dn_{ads}$(mmol/g) | $n_{ads}$ (wt%) | $dn_{ads}$(wt%) |
|---|---|---|---|---|---|---|
| 5.00E-05 | 0.9793 | 0.9799 | 2.92E-03 | 2.92E-03 | 5.88E-04 | 5.89E-04 |
| 1.00E-04 | 1.9681 | 1.3941 | 5.87E-03 | 4.16E-03 | 1.18E-03 | 8.38E-04 |
| 5.00E-04 | 9.6941 | 3.1072 | 2.89E-02 | 9.26E-03 | 5.82E-03 | 1.87E-03 |
| 1.00E-03 | 18.2742 | 4.2214 | 5.45E-02 | 1.26E-02 | 1.10E-02 | 2.54E-03 |
| 2.00E-03 | 37.0337 | 5.8868 | 0.110415 | 1.76E-02 | 2.22E-02 | 3.54E-03 |
| 5.00E-03 | 86.564 | 9.1233 | 0.258089 | 2.72E-02 | 5.20E-02 | 5.48E-03 |
| 1.00E-02 | 159.1712 | 10.9888 | 0.474567 | 3.28E-02 | 9.55E-02 | 6.59E-03 |
| 2.00E-02 | 292.1339 | 15.2986 | 0.870993 | 4.56E-02 | 0.175224 | 9.16E-03 |
| 3.00E-02 | 403.2446 | 18.103 | 1.202268 | 5.40E-02 | 0.241707 | 1.08E-02 |
| 5.00E-02 | 615.1374 | 20.7719 | 1.834023 | 6.19E-02 | 0.368249 | 1.24E-02 |
| 7.00E-02 | 792.3472 | 25.3684 | 2.362372 | 7.56E-02 | 0.473833 | 1.51E-02 |
| 0.1 | 1028.98 | 24.7252 | 3.06789 | 7.37E-02 | 0.614472 | 1.47E-02 |
| 0.2 | 1669.105 | 32.4291 | 4.976411 | 9.67E-02 | 0.992937 | 1.91E-02 |
| 0.3 | 2128.982 | 35.0506 | 6.34753 | 0.104503 | 1.26306 | 2.05E-02 |
| 0.5 | 2785.79 | 32.8724 | 8.305792 | 9.80E-02 | 1.646308 | 1.91E-02 |
| 0.7 | 3266.283 | 38.2574 | 9.738376 | 0.114064 | 1.924798 | 2.21E-02 |
| 1 | 3793.544 | 33.9034 | 11.3104 | 0.101083 | 2.228585 | 1.95E-02 |
| 2 | 4766.122 | 31.5905 | 14.21013 | 9.42E-02 | 2.784036 | 1.79E-02 |
| 3 | 5302.112 | 26.5291 | 15.80817 | 7.91E-02 | 3.087457 | 1.50E-02 |
| 5 | 5871.041 | 28.9583 | 17.50442 | 8.63E-02 | 3.40746 | 1.62E-02 |
| 7 | 6398.251 | 45.2376 | 19.07629 | 0.134875 | 3.702117 | 2.52E-02 |
| 10 | 6808.344 | 32.9688 | 20.29898 | 9.83E-02 | 3.930077 | 1.83E-02 |

Table S9. Adsorption isotherm of DL hydrogen in hierarchical silicalite HYZ-NH at 298 K.

| Pressure (katm) | N (molecules) | dN (molecules) | $n_{ads}$(mmol/g) | $dn_{ads}$(mmol/g) | $n_{ads}$ (wt%) | $dn_{ads}$(wt%) |
|---|---|---|---|---|---|---|
| 5.00E-05 | 1.7995 | 1.1821 | 5.39E-03 | 3.54E-03 | 1.09E-03 | 7.14E-04 |
| 1.00E-04 | 3.4494 | 1.6136 | 1.03E-02 | 4.83E-03 | 2.08E-03 | 9.74E-04 |
| 5.00E-04 | 12.0163 | 3.2981 | 3.60E-02 | 9.88E-03 | 7.25E-03 | 1.99E-03 |
| 1.00E-03 | 21.3649 | 4.1704 | 6.40E-02 | 1.25E-02 | 1.29E-02 | 2.52E-03 |
| 2.00E-03 | 40.5669 | 6.2026 | 0.12152 | 1.86E-02 | 2.45E-02 | 3.74E-03 |
| 5.00E-03 | 91.4211 | 9.038 | 0.273857 | 2.71E-02 | 5.52E-02 | 5.45E-03 |
| 1.00E-02 | 161.4152 | 11.6837 | 0.483528 | 3.50E-02 | 9.74E-02 | 7.04E-03 |
| 2.00E-02 | 296.2732 | 16.5394 | 0.887503 | 4.95E-02 | 0.178539 | 9.95E-03 |
| 3.00E-02 | 412.1112 | 18.2785 | 1.234502 | 5.48E-02 | 0.248172 | 1.10E-02 |
| 5.00E-02 | 618.0092 | 21.974 | 1.851282 | 6.58E-02 | 0.371702 | 1.32E-02 |
| 7.00E-02 | 802.9007 | 24.8313 | 2.405134 | 7.44E-02 | 0.482368 | 1.48E-02 |
| 0.1 | 1034.762 | 27.2252 | 3.099688 | 8.16E-02 | 0.620802 | 1.62E-02 |
| 0.2 | 1669.152 | 29.0024 | 5.000038 | 8.69E-02 | 0.997605 | 1.72E-02 |
| 0.3 | 2130.167 | 34.6916 | 6.381035 | 0.103921 | 1.269642 | 2.04E-02 |
| 0.5 | 2808.221 | 37.1073 | 8.412186 | 0.111157 | 1.667045 | 2.17E-02 |
| 0.7 | 3286.867 | 42.6379 | 9.845996 | 0.127724 | 1.945655 | 2.47E-02 |
| 1 | 3785.464 | 36.3237 | 11.33957 | 0.10881 | 2.234205 | 2.10E-02 |
| 2 | 4802.943 | 31.2506 | 14.38749 | 9.36E-02 | 2.817806 | 1.78E-02 |
| 3 | 5343.531 | 35.052 | 16.00685 | 0.105 | 3.125048 | 1.99E-02 |
| 5 | 5995.941 | 33.0142 | 17.96118 | 9.89E-02 | 3.493268 | 1.86E-02 |
| 7 | 6451.97 | 31.2547 | 19.32724 | 9.36E-02 | 3.748992 | 1.75E-02 |
| 10 | 6873.703 | 32.1548 | 20.59056 | 9.63E-02 | 3.984281 | 1.79E-02 |

Table S10. Adsorption isotherm of Buch hydrogen in hierarchical silicalite HYY-OH at 298 K.

| Pressure (katm) | N (molecules) | dN (molecules) | $n_{ads}$(mmol/g) | $dn_{ads}$(mmol/g) | $n_{ads}$ (wt%) | $dn_{ads}$(wt%) |
|---|---|---|---|---|---|---|
| 5.00E-05 | 0.511 | 0.7132 | 1.44E-03 | 2.01E-03 | 2.90E-04 | 4.05E-04 |
| 1.00E-04 | 1.0186 | 1.0053 | 2.87E-03 | 2.83E-03 | 5.79E-04 | 5.71E-04 |
| 5.00E-04 | 5.0463 | 2.2573 | 1.42E-02 | 6.36E-03 | 2.87E-03 | 1.28E-03 |
| 1.00E-03 | 10.0921 | 3.1897 | 2.85E-02 | 8.99E-03 | 5.73E-03 | 1.81E-03 |
| 2.00E-03 | 20.2094 | 4.5574 | 5.70E-02 | 1.28E-02 | 1.15E-02 | 2.59E-03 |
| 5.00E-03 | 49.9427 | 6.9911 | 0.140807 | 1.97E-02 | 2.84E-02 | 3.97E-03 |
| 1.00E-02 | 97.9518 | 9.8227 | 0.276163 | 2.77E-02 | 5.56E-02 | 5.57E-03 |
| 2.00E-02 | 191.9598 | 13.9651 | 0.541208 | 3.94E-02 | 0.108951 | 7.92E-03 |
| 3.00E-02 | 280.2799 | 15.917 | 0.790216 | 4.49E-02 | 0.158999 | 9.02E-03 |
| 5.00E-02 | 446.6871 | 21.1198 | 1.259381 | 5.95E-02 | 0.25316 | 1.19E-02 |
| 7.00E-02 | 594.0951 | 22.7673 | 1.674981 | 6.42E-02 | 0.336423 | 1.28E-02 |
| 0.1 | 806.4948 | 25.7675 | 2.273816 | 7.26E-02 | 0.456152 | 1.45E-02 |
| 0.2 | 1364.76 | 30.6195 | 3.847777 | 8.63E-02 | 0.769475 | 1.71E-02 |
| 0.3 | 1782.449 | 34.3325 | 5.025402 | 9.68E-02 | 1.002614 | 1.91E-02 |
| 0.5 | 2415.861 | 36.5946 | 6.811232 | 0.103174 | 1.354079 | 2.02E-02 |
| 0.7 | 2879.666 | 42.4401 | 8.118875 | 0.119655 | 1.609855 | 2.33E-02 |
| 1 | 3366.317 | 35.2821 | 9.49093 | 9.95E-02 | 1.876808 | 1.93E-02 |
| 2 | 4343.478 | 44.3858 | 12.24592 | 0.12514 | 2.408479 | 2.40E-02 |
| 3 | 4870.483 | 37.5094 | 13.73175 | 0.105753 | 2.692836 | 2.02E-02 |
| 5 | 5547.158 | 37.2158 | 15.63955 | 0.104926 | 3.055532 | 1.99E-02 |
| 7 | 5983.966 | 35.2314 | 16.87108 | 9.93E-02 | 3.288225 | 1.87E-02 |
| 10 | 6393.519 | 24.7368 | 18.02576 | 6.97E-02 | 3.505388 | 1.31E-02 |

Table S11. Adsorption isotherm of DL hydrogen in hierarchical silicalite HYY-OH at 298 K.

| Pressure (katm) | N (molecules) | dN (molecules) | $n_{ads}$(mmol/g) | $dn_{ads}$(mmol/g) | $n_{ads}$ (wt%) | $dn_{ads}$(wt%) |
|---|---|---|---|---|---|---|
| 5.00E-05 | 5.3867 | 1.1267 | 1.52E-02 | 3.18E-03 | 3.06E-03 | 6.40E-04 |
| 1.00E-04 | 10.638 | 1.3215 | 3.00E-02 | 3.73E-03 | 6.04E-03 | 7.51E-04 |
| 5.00E-04 | 16.0221 | 2.7087 | 4.52E-02 | 7.64E-03 | 9.10E-03 | 1.54E-03 |
| 1.00E-03 | 25.9554 | 3.7417 | 7.32E-02 | 1.05E-02 | 1.47E-02 | 2.13E-03 |
| 2.00E-03 | 40.9159 | 5.6562 | 0.115358 | 1.59E-02 | 2.32E-02 | 3.21E-03 |
| 5.00E-03 | 80.8658 | 8.1097 | 0.227992 | 2.29E-02 | 4.59E-02 | 4.60E-03 |
| 1.00E-02 | 139.3447 | 10.9022 | 0.392866 | 3.07E-02 | 7.91E-02 | 6.18E-03 |
| 2.00E-02 | 250.8791 | 15.4384 | 0.707324 | 4.35E-02 | 0.142344 | 8.75E-03 |
| 3.00E-02 | 355.3631 | 17.4455 | 1.001904 | 4.92E-02 | 0.201507 | 9.87E-03 |
| 5.00E-02 | 549.6475 | 21.9384 | 1.549666 | 6.19E-02 | 0.311332 | 1.24E-02 |
| 7.00E-02 | 719.0818 | 26.0026 | 2.027366 | 7.33E-02 | 0.406912 | 1.47E-02 |
| 0.1 | 947.3923 | 28.8795 | 2.67106 | 8.14E-02 | 0.535416 | 1.62E-02 |
| 0.2 | 1544.103 | 34.4502 | 4.353415 | 9.71E-02 | 0.869713 | 1.92E-02 |
| 0.3 | 1994.246 | 41.4756 | 5.622539 | 0.116936 | 1.120414 | 2.30E-02 |
| 0.5 | 2636.952 | 40.8436 | 7.434572 | 0.115154 | 1.476171 | 2.25E-02 |
| 0.7 | 3099.354 | 37.1492 | 8.738258 | 0.104738 | 1.730545 | 2.04E-02 |
| 1 | 3592.228 | 41.2029 | 10.12786 | 0.116167 | 2.00024 | 2.25E-02 |
| 2 | 4575.595 | 27.4215 | 12.90034 | 7.73E-02 | 2.533927 | 1.48E-02 |
| 3 | 5105.221 | 32.5618 | 14.39356 | 9.18E-02 | 2.818962 | 1.75E-02 |
| 5 | 5753.482 | 39.8003 | 16.22126 | 0.112212 | 3.165583 | 2.12E-02 |
| 7 | 6187.006 | 39.7018 | 17.44353 | 0.111935 | 3.396008 | 2.11E-02 |
| 10 | 6602.777 | 26.0205 | 18.61574 | 7.34E-02 | 3.615971 | 1.37E-02 |

Table S12. Adsorption isotherm of DLnc hydrogen in hierarchical silicalite HYY-OH at 298 K.

| Pressure (katm) | N (molecules) | dN (molecules) | $n_{ads}$(mmol/g) | $dn_{ads}$(mmol/g) | $n_{ads}$ (wt%) | $dn_{ads}$(wt%) |
|---|---|---|---|---|---|---|
| 5.00E-05 | 0.5256 | 0.7258 | 1.48E-03 | 2.05E-03 | 2.99E-04 | 4.12E-04 |
| 1.00E-04 | 1.0563 | 1.0268 | 2.98E-03 | 2.89E-03 | 6.00E-04 | 5.83E-04 |
| 5.00E-04 | 5.277 | 2.2911 | 1.49E-02 | 6.46E-03 | 3.00E-03 | 1.30E-03 |
| 1.00E-03 | 10.5308 | 3.2684 | 2.97E-02 | 9.21E-03 | 5.98E-03 | 1.86E-03 |
| 2.00E-03 | 20.9759 | 4.5519 | 5.91E-02 | 1.28E-02 | 1.19E-02 | 2.59E-03 |
| 5.00E-03 | 51.8124 | 7.2313 | 0.146079 | 2.04E-02 | 2.94E-02 | 4.11E-03 |
| 1.00E-02 | 102.0629 | 10.4809 | 0.287754 | 2.95E-02 | 5.80E-02 | 5.95E-03 |
| 2.00E-02 | 198.3387 | 14.3156 | 0.559192 | 4.04E-02 | 0.112567 | 8.12E-03 |
| 3.00E-02 | 289.1457 | 17.8202 | 0.815212 | 5.02E-02 | 0.16402 | 1.01E-02 |
| 5.00E-02 | 461.4705 | 21.0628 | 1.301061 | 5.94E-02 | 0.261517 | 1.19E-02 |
| 7.00E-02 | 614.7532 | 25.0016 | 1.733223 | 7.05E-02 | 0.34808 | 1.41E-02 |
| 0.1 | 824.848 | 26.861 | 2.325561 | 7.57E-02 | 0.466484 | 1.51E-02 |
| 0.2 | 1396.613 | 34.1671 | 3.937583 | 9.63E-02 | 0.787293 | 1.91E-02 |
| 0.3 | 1812.446 | 30.6013 | 5.109975 | 8.63E-02 | 1.019315 | 1.70E-02 |
| 0.5 | 2447.442 | 37.4899 | 6.900271 | 0.105698 | 1.371538 | 2.07E-02 |
| 0.7 | 2917.905 | 36.2613 | 8.226686 | 0.102234 | 1.630884 | 1.99E-02 |
| 1 | 3406.713 | 37.1193 | 9.604823 | 0.104653 | 1.898902 | 2.03E-02 |
| 2 | 4370.907 | 40.0131 | 12.32325 | 0.112812 | 2.42332 | 2.16E-02 |
| 3 | 4912.497 | 29.2362 | 13.8502 | 8.24E-02 | 2.715435 | 1.57E-02 |
| 5 | 5598.826 | 43.1437 | 15.78522 | 0.121639 | 3.083114 | 2.30E-02 |
| 7 | 5979.165 | 35.7592 | 16.85754 | 0.100819 | 3.285674 | 1.90E-02 |
| 10 | 6402.925 | 33.0257 | 18.05229 | 9.31E-02 | 3.510365 | 1.75E-02 |

3. Temperature Variation of adsorption amounts

The following tables list the temperature variation of adsorption amounts of hydrogen in the adsorbent models IC and HZZ-OH using different force fields. The first column shows the simulation temperature. The second and third columns show respectively the adsorbed amounts in wt% at 0.1 atm and the corresponding standard deviation. The fourth and fifth columns show these quantities at 1 atm, while the sixth and seventh columns those at 10 atm. The last two columns (eight and ninth) show the amounts of hydrogen adsorbed at 100 atm and the corresponding standard deviation.

Table S13. Temperature variation of adsorbed amounts of Buch hydrogen at four representative pressures in silicalite IC.

| Temperature (K) | $n_{ads}$ \| P= 0.1 atm (wt%) | $dn_{ads}$ \| P =0.1 atm (wt%) | $n_{ads}$ \| P= 1 atm (wt%) | $dn_{ads}$ \| P =1 atm (wt%) | $n_{ads}$ \| P= 10 atm (wt%) | $dn_{ads}$ \| P =10 atm (wt%) | $n_{ads}$ \| P= 100 atm (wt%) | $dn_{ads}$ \| P= 100 atm (wt%) |
|---|---|---|---|---|---|---|---|---|
| 273.15 | 3.86E-04 | 3.77E-04 | 3.84E-03 | 1.19E-03 | 3.67E-02 | 3.72E-03 | 0.251159 | 8.70E-03 |
| 278.15 | 3.65E-04 | 3.65E-04 | 3.62E-03 | 1.15E-03 | 3.49E-02 | 3.57E-03 | 0.242097 | 9.60E-03 |
| 283.15 | 3.46E-04 | 3.56E-04 | 3.44E-03 | 1.12E-03 | 3.30E-02 | 3.44E-03 | 0.234998 | 9.12E-03 |
| 288.15 | 3.34E-04 | 3.49E-04 | 3.28E-03 | 1.09E-03 | 3.14E-02 | 3.47E-03 | 0.225825 | 8.66E-03 |
| 293.15 | 3.16E-04 | 3.42E-04 | 3.13E-03 | 1.07E-03 | 3.01E-02 | 3.35E-03 | 0.219119 | 8.87E-03 |
| 298.15 | 3.04E-04 | 3.31E-04 | 2.97E-03 | 1.05E-03 | 2.89E-02 | 3.23E-03 | 0.211823 | 8.18E-03 |
| 303.15 | 2.92E-04 | 3.26E-04 | 2.87E-03 | 1.02E-03 | 2.76E-02 | 3.15E-03 | 0.20398 | 8.82E-03 |
| 308.15 | 2.74E-04 | 3.17E-04 | 2.74E-03 | 1.00E-03 | 2.65E-02 | 3.09E-03 | 0.196913 | 8.02E-03 |

Table S14. Temperature variation of adsorbed amounts of DL hydrogen at four representative pressures in silicalite IC.

| Temperature (K) | $n_{ads}$ \| P= 0.1 atm (wt%) | $dn_{ads}$ \| P =0.1 atm (wt%) | $n_{ads}$ \| P= 1 atm (wt%) | $dn_{ads}$ \| P =1 atm (wt%) | $n_{ads}$ \| P= 10 atm (wt%) | $dn_{ads}$ \| P =10 atm (wt%) | $n_{ads}$ \| P= 100 atm (wt%) | $dn_{ads}$ \| P= 100 atm (wt%) |
|---|---|---|---|---|---|---|---|---|
| 273.15 | 4.40E-04 | 4.00E-04 | 4.32E-03 | 1.25E-03 | 4.10E-02 | 3.89E-03 | 0.272851 | 7.88E-03 |
| 278.15 | 4.17E-04 | 3.91E-04 | 4.06E-03 | 1.21E-03 | 3.87E-02 | 3.75E-03 | 0.264262 | 8.25E-03 |
| 283.15 | 3.88E-04 | 3.74E-04 | 3.86E-03 | 1.19E-03 | 3.70E-02 | 3.64E-03 | 0.254852 | 7.54E-03 |
| 288.15 | 3.68E-04 | 3.66E-04 | 3.70E-03 | 1.16E-03 | 3.50E-02 | 3.57E-03 | 0.244551 | 7.90E-03 |
| 293.15 | 3.53E-04 | 3.59E-04 | 3.48E-03 | 1.12E-03 | 3.34E-02 | 3.51E-03 | 0.236514 | 8.98E-03 |
| 298.15 | 3.36E-04 | 3.49E-04 | 3.31E-03 | 1.11E-03 | 3.18E-02 | 3.34E-03 | 0.22856 | 8.07E-03 |
| 303.15 | 3.18E-04 | 3.39E-04 | 3.17E-03 | 1.07E-03 | 3.04E-02 | 3.34E-03 | 0.22075 | 7.43E-03 |
| 308.15 | 3.02E-04 | 3.33E-04 | 3.01E-03 | 1.04E-03 | 2.92E-02 | 3.23E-03 | 0.214789 | 7.97E-03 |

Table S15. Temperature variation of adsorbed amounts of Buch hydrogen at four representative pressures in hierarchical silicalite HZZ-OH.

| Temperature (K) | $n_{ads}$ I P= 0.1 atm (wt%) | $dn_{ads}$ I P =0.1 atm (wt%) | $n_{ads}$ I P= 1 atm (wt%) | $dn_{ads}$ I P =1 atm (wt%) | $n_{ads}$ I P= 10 atm (wt%) | $dn_{ads}$ I P =10 atm (wt%) | $n_{ads}$ I P= 100 atm (wt%) | $dn_{ads}$ I P= 100 atm (wt%) |
|---|---|---|---|---|---|---|---|---|
| 273.15 | 6.93E-04 | 6.29E-04 | 6.90E-03 | 1.98E-03 | 6.66E-02 | 6.01E-03 | 0.517992 | 1.62E-02 |
| 278.15 | 6.67E-04 | 6.14E-04 | 6.67E-03 | 1.97E-03 | 6.40E-02 | 5.94E-03 | 0.505305 | 1.51E-02 |
| 283.15 | 6.47E-04 | 6.09E-04 | 6.38E-03 | 1.91E-03 | 6.21E-02 | 5.88E-03 | 0.492205 | 1.45E-02 |
| 288.15 | 6.21E-04 | 5.91E-04 | 6.20E-03 | 1.87E-03 | 5.95E-02 | 5.93E-03 | 0.477491 | 1.54E-02 |
| 293.15 | 6.01E-04 | 5.85E-04 | 6.00E-03 | 1.86E-03 | 5.80E-02 | 5.71E-03 | 0.466809 | 1.52E-02 |
| 298.15 | 5.79E-04 | 5.71E-04 | 5.73E-03 | 1.81E-03 | 5.56E-02 | 5.57E-03 | 0.456152 | 1.45E-02 |
| 303.15 | 5.60E-04 | 5.63E-04 | 5.59E-03 | 1.79E-03 | 5.39E-02 | 5.41E-03 | 0.443002 | 1.56E-02 |
| 308.15 | 5.41E-04 | 5.54E-04 | 5.41E-03 | 1.76E-03 | 5.26E-02 | 5.42E-03 | 0.432064 | 1.43E-02 |

Table S16. Temperature variation of adsorbed amounts of DL hydrogen at four representative pressures in hierarchical silicalite HZZ-OH.

| Temperature (K) | $n_{ads}$ I P= 0.1 atm (wt%) | $dn_{ads}$ I P =0.1 atm (wt%) | $n_{ads}$ I P= 1 atm (wt%) | $dn_{ads}$ I P =1 atm (wt%) | $n_{ads}$ I P= 10 atm (wt%) | $dn_{ads}$ I P =10 atm (wt%) | $n_{ads}$ I P= 100 atm (wt%) | $dn_{ads}$ I P= 100 atm (wt%) |
|---|---|---|---|---|---|---|---|---|
| 273.15 | 4.41E-03 | 9.01E-04 | 1.82E-02 | 2.33E-03 | 9.65E-02 | 7.28E-03 | 0.618038 | 1.80E-02 |
| 278.15 | 6.66E-03 | 7.76E-04 | 1.83E-02 | 2.31E-03 | 9.35E-02 | 6.91E-03 | 0.603025 | 1.65E-02 |
| 283.15 | 6.64E-03 | 7.65E-04 | 1.67E-02 | 2.34E-03 | 8.89E-02 | 6.45E-03 | 0.580063 | 1.77E-02 |
| 288.15 | 4.82E-03 | 8.12E-04 | 1.60E-02 | 2.16E-03 | 8.46E-02 | 6.65E-03 | 0.561907 | 1.60E-02 |
| 293.15 | 5.51E-03 | 7.15E-04 | 1.51E-02 | 2.12E-03 | 8.10E-02 | 6.27E-03 | 0.550022 | 1.57E-02 |
| 298.15 | 6.04E-03 | 6.40E-04 | 1.47E-02 | 2.13E-03 | 7.91E-02 | 6.18E-03 | 0.535416 | 1.62E-02 |
| 303.15 | 4.22E-03 | 6.69E-04 | 1.41E-02 | 2.05E-03 | 7.54E-02 | 6.43E-03 | 0.518438 | 1.66E-02 |
| 308.15 | 3.54E-03 | 6.33E-04 | 1.39E-02 | 2.01E-03 | 7.32E-02 | 6.03E-03 | 0.50252 | 1.67E-02 |